\documentclass[runningheads]{llncs}

\usepackage[T1]{fontenc}
\usepackage{graphicx}
\usepackage{tikz}
\usepackage{booktabs}
\usepackage{wrapfig}
\usepackage{amsmath}
\usepackage{amssymb}
\usepackage{algorithm}
\usepackage{algpseudocode}
\usepackage{array}
\usepackage[hyphens]{url}
\usepackage{multirow}
\usepackage{xcolor}
\usepackage{needspace}
\usetikzlibrary{arrows.meta,positioning}

\title{Conditions for Social Trajectory Collapse:\\ Agent-Based Simulation of Time-Geographic Trajectory Distributions}
\titlerunning{Conditions for Social Trajectory Collapse}

\author{Daneul Kim\inst{1}\textsuperscript{*,\ensuremath{\dagger}}
\and
Yuyeong Kim\inst{2}\textsuperscript{*}}
\authorrunning{D. Kim and Y. Kim}

\institute{
Seoul National University, Republic of Korea
\and
NC AI, Republic of Korea
}

\begin{document}

\maketitle

\begingroup
\renewcommand{\thefootnote}{{*}}
\footnotetext{Equal contribution.}
\renewcommand{\thefootnote}{{\ensuremath{\dagger}}}
\footnotetext{Corresponding author.}
\endgroup
\setcounter{footnote}{0}

\begin{abstract}
Metropolitan concentration has motivated policies for more balanced regional development.
We use agent-based simulation to examine trajectory diversity, the variety of people's recurring activity orientations toward neighborhoods, urban centers, and wider networks.
The model maps 2010 population shares to 2026 distributions in five country cases, China, Russia, Japan, the United Kingdom, and the United States, and simulates change to 2036.
Trajectory diversity declines by 52.6--79.0\% over 2026--2036.
Russia and Japan show the largest increases in the share of the most common trajectory, while US diversity contracts despite a slight decline in this share.
Cost burdens rise throughout, whereas welfare falls in the Russian and Japanese cases but rises in the other three.
These findings suggest that regional development should be assessed through activity diversity, cost burdens, and welfare together.
\keywords{Computational social science \and Agent-based modeling \and Accessibility \and Time-geographic trajectories \and Urban concentration}
\end{abstract}

\Needspace{6\baselineskip}
\section{Introduction}

Concentration of population and opportunity in major cities has made balanced regional development a recurring policy concern.
South Korea's Innovation Cities initiative relocated public institutions from the capital region to support regional centers \cite{ahn2021regional}.
The United Kingdom has announced further relocation of civil service roles from London to support regional growth \cite{cabinetoffice2025relocation}.
These efforts motivate our question: how does everyday activity become concentrated around particular cities and districts, and how might this change over the next decade?

We use \emph{trajectory} to describe the spatial reach and orientation of a person's recurring daily activities, inspired by time geography \cite{hagerstrand1970people,ellegard2012timegeography}.
A trajectory connects the local neighborhood with primary urban centers, regional centers, and wider networks.
\emph{Trajectory diversity} is the variety of these activity orientations across the population.
For example, one person may organize activity around nearby services and local ties, while another relies on urban centers and wider networks.
Diversity reflects how many such orientations coexist and how evenly they are represented.
We call a decline in this diversity \emph{social trajectory collapse}.
Activity orientations and regional population shares describe complementary aspects of concentration and are measured separately in the model.

Better transportation, communication, and information access can expand the opportunities people reach.
When opportunity and visibility concentrate in particular cities or districts, improved access may also draw activity toward those centers.
Rising housing and service costs can constrain access, while regional support may sustain activity around other centers.
These pressures make it difficult to infer changes in activity diversity from improvements in connectivity alone.
We ask how much trajectory diversity changes over a decade and how those changes relate to cost burdens and welfare.

We develop an agent-based model in which agents update their activity orientations using regional conditions, local peers, and a population activity signal, while demand feeds back into regional costs.
The model links individual responses to these conditions with changes in the population distribution of activity orientations.
We examine five country cases: China, Russia, Japan, the United Kingdom (UK), and the United States (US).
Starting from 2010 population shares, we fit each case's 2026 distribution, check annual 2020--2025 shares, and simulate 2026--2036 using the selected settings.
Diversity declines in all five cases, while welfare changes differ.
Russia and Japan show the strongest increases in the share of the most common trajectory.
The US loses diversity despite a small decline in this share, illustrating the importance of examining the full distribution.
Our contribution is to connect urban concentration to a common measure of recurring activity diversity and examine it alongside cost and welfare.
Combining diversity with the share of the most common trajectory reveals different forms of concentration across the cases.
We assess these findings across simulation trials, alternative grouping rules, parameter perturbations, and agent counts.

\section{Related Work}

Agent-based modeling explains aggregate patterns through individual decisions and interactions \cite{epstein2006generative,schelling1971dynamic}.
Social influence can increase inequality in collective outcomes \cite{salganik2006experimental}, while recommender feedback can homogenize user behavior \cite{chaney2018algorithmic}.
These mechanisms motivate our treatment of peer activity and platform visibility as influences on recurring spatial activity.
The observed regularity of individual mobility supports examining how recurring activity is distributed across a population \cite{gonzalez2008mobility,song2010predictability}.

Accessibility research treats access as multidimensional, incorporating transport and information rather than travel speed alone \cite{geurs2004accessibility,chorus2013ict}.
The capability approach and spatial-opportunity research distinguish available opportunities from those people can access \cite{sen1992inequality,robeyns2005capability,galster2017spatial}.
Migration research similarly separates aspirations from the capacity to move, while social media shapes knowledge of opportunities elsewhere \cite{dehaas2021migration,dekker2014socialmedia}.
Together, these perspectives motivate distinguishing regional opportunity from agents' ability to reach and use it.

Urban economics links concentration to increasing returns, rent capitalization, and housing supply \cite{krugman1991increasing,roback1982wages,glaeser2005urban}.
Polycentric governance coordinates collective action across multiple scales \cite{ostrom2010polycentric}.
Our model connects these spatial mechanisms with social influence to examine how activity diversity evolves as regional costs respond to demand.
Measuring diversity, cost, and welfare together allows us to compare how concentration accompanies changes in daily life across urban systems.

\section{Method}

\subsection{Model Overview}

We model social trajectory collapse as declining diversity in city-oriented activity trajectories, drawing on ODD's distinction between entities, state variables, and submodels \cite{grimm2010odd}.
Agent $i$ has a six-dimensional state
\begin{equation}
x_i(t)=(L_i,P_i,R_i,N_i,C_i,S_i)\in[0,1]^6,
\end{equation}
where $L$ is local embeddedness, $P$ primary-center access, $R$ regional-center access, $N$ network access, $C$ cost strain, and $S$ support security.
The activity orientation is $z_i=(L_i,P_i,R_i,N_i)$, while current region identity is recorded separately.
Regions have opportunity, cost, support, and visibility features, while connectivity and housing-supply elasticity are case-level settings.

Population-based regional fields initialize six features corresponding to the agent-state coordinates.
Each annual step updates states using current regional features, local peer means, and the population activity signal, modified by exposure, individual inertia and sensitivity, cost and support adjustments, and stochastic variation.
Connectivity and amplification modify network and primary-center exposure.
For candidate region $a$, the movement score is $U_{i,a}=z_i^{\mathsf T}f_{a,1:4}-w_C C_a+w_S S_a$, where $f_{a,1:4}$ contains the first four regional features and $w_C,w_S$ are fitted weights.
Candidate regions are sampled using
\begin{equation}
P(a_i(t)=a)=\frac{\exp(U_{i,a}(t)/\tau)}
{\sum_{a'}\exp(U_{i,a'}(t)/\tau)},
\end{equation}
over all regions, including the current region, with a separate connectivity-dependent migration probability gating adult moves.
Activity orientations can therefore change even when an agent remains in the same region.
Regional density and aggregate primary-center/network activity raise regional cost, moderated by housing-supply elasticity, feeding into the next movement update.
Density is measured relative to initial regional population.
Perception and state updating precede region choice, mobility, cost feedback, and measurement in each annual step.

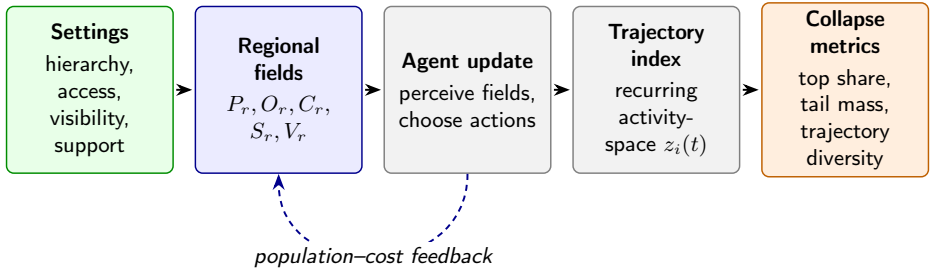
\begin{figure}[t]
\begin{center}
\resizebox{\linewidth}{!}{%
\begin{tikzpicture}[
  >=Stealth,
  font=\sffamily\small,
  node distance=3mm,
  box/.style={
    draw, semithick, rounded corners=3pt,
    align=center,
    text width=20mm,
    minimum width=24mm,
    minimum height=24mm,
    inner xsep=3pt, inner ysep=3pt
  },
  input/.style  ={box, draw=green!55!black,  fill=green!9},
  field/.style  ={box, draw=blue!55!black,   fill=blue!8},
  process/.style={box, draw=black!50,        fill=black!5},
  measure/.style={box, draw=orange!75!black, fill=orange!13},
  flow/.style={->, thick, shorten >=1pt, shorten <=1pt},
  feed/.style={->, thick, draw=blue!55!black,
               dash pattern=on 2.5pt off 2pt}
]
\node[input]   (settings)                    {\textbf{Settings}\\[2pt]
  hierarchy, access, visibility, support};
\node[field]   (fields)  [right=of settings] {\textbf{Regional fields}\\[2pt]
  $P_r,O_r,C_r,$\\ $S_r,V_r$};
\node[process] (agents)  [right=of fields]   {\textbf{Agent update}\\[2pt]
  perceive fields, choose actions};
\node[process] (traj)    [right=of agents]   {\textbf{Trajectory index}\\[2pt]
  recurring activity-space $z_i(t)$};
\node[measure] (metrics) [right=of traj]     {\textbf{Collapse metrics}\\[2pt]
  top share, tail mass, trajectory diversity};

\foreach \a/\b in {settings/fields, fields/agents, agents/traj, traj/metrics}
  \draw[flow] (\a) -- (\b);

\draw[feed] (agents.south) to[out=-90, in=-90, looseness=1.5]
  node[midway, fill=white, inner sep=2pt,
       font=\sffamily\footnotesize\itshape]
  {population--cost feedback}
  (fields.south);
\end{tikzpicture}%
}
\end{center}
\vspace{-4mm}
\caption{Model overview.
Regional fields and social signals update agents' activity orientations, while demand feeds back into regional cost.
Similar orientations are grouped to measure trajectory diversity and concentration.}
\label{fig:concept-overview}
\end{figure}

For measurement, we group agents by similar values of $z_i$ using the same rule across all cases.
Agents living in different cities can belong to the same group if their activity orientations are similar.
Conversely, residents of one city can occupy different groups, allowing activity diversity and population concentration to vary separately.
Regional cost and support enter movement and welfare calculations.
For population share $\pi_c$ in group $c$, the summary measures are:
\begin{equation}
q_{\mathrm{top}}=\max_c\pi_c,\quad
m_{\mathrm{tail}}=1-q_{\mathrm{top}},\quad
H=-\sum_c\pi_c\log\pi_c,\quad
N_{\mathrm{eff}}=\exp(H).
\end{equation}
Here $q_{\mathrm{top}}$ is the share of the most common trajectory, $m_{\mathrm{tail}}$ the remaining share, and $H$ entropy.
Trajectory diversity $N_{\mathrm{eff}}$ is the effective number of groups, reflecting both their number and the evenness of their population shares.
For an equal distribution across $k$ groups, $N_{\mathrm{eff}}=k$, giving the measure an intuitive group-count interpretation.
It can decline even when $q_{\mathrm{top}}$ falls, capturing contraction across the full distribution.
Cost burden is mean current regional cost, $\overline C=n^{-1}\sum_i C_{r_i}$, where $r_i$ is agent $i$'s region and $n$ the fixed population size.
Welfare combines activity-state coordinates with current regional cost and support, as specified below.
Changes subtract each case's initial value and are interpreted within that case.

\subsection{Region Generator}

Official population shares $P_r$, hierarchy scores, hub bonuses, and supply profiles generate bounded regional features through linear, power/scaling, or hinge families.
Opportunity depends on population, hierarchy, and hubs, while visibility depends on opportunity and population.
Cost combines population, opportunity, visibility, and supply, with inputs varying by family.
These model families draw on scaling and capitalization research \cite{bettencourt2007scaling,duranton2004micro,roback1982wages,glaeser2005urban} and adapt social-influence and threshold mechanisms to visibility \cite{salganik2006experimental,granovetter1978threshold}.
Motivated by local public goods and polycentric governance \cite{tiebout1956local,ostrom2010polycentric}, we generate support by mixing a uniform component with a population-weighted opportunity/cost proxy for fiscal capacity, followed by rescaling.

\subsection{Real-World Settings}

We evaluate five selected urban systems using World Urbanization Prospects (WUP) 2010 population estimates and 2026 projections \cite{unitednations2025wup}.
Country labels denote selected major urban regions in China, the UK, and the US, and regional networks centered on Moscow and Tokyo.
The China case comprises 14 urban regions, including Shanghai, Beijing, Hong Kong, Shenzhen, and Guangzhou.
The US case comprises 17 urban regions, including New York, Los Angeles, Chicago, and Washington, DC.
The UK case comprises 12 urban regions, including London, Birmingham, Manchester, Edinburgh, Cardiff, and Belfast.
The Russia case comprises Moscow, Tver, Kaluga, Tula, Ryazan, Vladimir, Yaroslavl, and Nizhny Novgorod.
The Japan case comprises Tokyo, Tsukuba, Utsunomiya, Takasaki--Maebashi, Mito, Shizuoka, and Nagano.
Population shares initialize the regional fields, and all listed cities are represented as separate regions.
The cases differ in geographic coverage and region count, which limits cross-case comparisons.

\subsection{Implementation Details}

Annual state updates use fixed weights of 0.48 for regional features, 0.24 for local peer means, and 0.28 for the population signal after exposure adjustments.
Trajectory measurement divides each coordinate of $z_i$ into six equal-width bins, defining $6^4=1{,}296$ groups and normalized entropy $H/\log(6^4)$.
Bin boundaries remain fixed throughout each simulation, so changes in group shares reflect evolving activity states under a common measurement rule.
Measurement sensitivity uses four and eight bins per coordinate, giving $4^4$ and $8^4$ possible groups, respectively.
These alternative grouping rules are applied to the same simulated states, separating measurement sensitivity from changes in the underlying dynamics.
Welfare is computed as
\begin{equation}
W=n^{-1}\sum_i[L_i+P_i+R_i+0.75N_i-1.15C_{r_i}+0.70S_{r_i}].
\end{equation}

Each simulation trial uses a different random seed to generate stochastic variation while keeping the model setting fixed.
For ensemble analyses, we run the five selected models with matching seed values and average their outputs within each trial.
The simulator uses PyTorch CUDA, with fixed total population and reproduction, child socialization, and policy interventions disabled.

\section{Experiments and Results}

One agent represents approximately 100,000 people in the main experiments.
Across the five cases, we search 1,215 global settings, refine locally, confirm candidates with 100 simulation trials, and retain an equal-weight ensemble of the five best settings per case.
We freeze the selected settings for all downstream experiments and initialize the main scenario from 2026 population shares, running 200 simulation trials to 2036.
The 2010--2026 mapping selects model settings, while the 2036 scenarios use a fresh initialization from WUP 2026 shares.

\subsection{2010--2026 Population Mapping}

We fit annual updates from 2010 shares by minimizing Hellinger distance between the trial-averaged simulated share vector and WUP 2026 shares.
Distances are computed after averaging share vectors across trials.
Table~\ref{tab:calibration-fit} compares the top-five ensemble and best single setting with persistence, which retains 2010 shares.
The ensemble improves on persistence in every case, although China's gain is small.
We report the best single setting for comparison and use the fixed top-five ensemble throughout the downstream analyses.
The best single setting has a slightly lower distance than the ensemble in four cases, while the US ensemble fits slightly better.

\begin{table*}[t]
\centering
\caption{2010--2026 population mapping, confirmed with 100 simulation trials.
Lower Hellinger distance is better.
Persistence retains 2010 shares.}
\label{tab:calibration-fit}
\small
\begin{tabular*}{\textwidth}{@{\extracolsep{\fill}}lrrr@{}}
\toprule
Case & Persistence & Top-five ensemble & Best single \\
\midrule
China  & 0.050175 & 0.049925 & 0.049841 \\
Russia       & 0.028824 & 0.021890 & 0.020194 \\
Japan        & 0.028644 & 0.014075 & 0.013705 \\
UK  & 0.018901 & 0.012588 & 0.012376 \\
US & 0.057516 & 0.052259 & 0.052310 \\
\bottomrule
\end{tabular*}
\end{table*}

With no intermediate-year retuning, 100 trials per model give equal-case-weight mean Hellinger distances rising from 0.028505 in 2020 to 0.030549 in 2025.
Hellinger, total variation, and Jensen--Shannon divergence attain their lowest cross-case means in 2020, while mean absolute error does so in 2022.
The mean Hellinger error rises gradually with increasing annual increments, forming a convex intermediate-year error profile.
This checks population paths within the fitting interval.

\subsection{2026--2036 Scenarios}

Table~\ref{tab:scenario-change} shows declining normalized entropy and trajectory diversity in every case, with diversity losses of 52.6--79.0\% over 2026--2036.
Russia and Japan show the largest increases in the share of the most common trajectory.
These increases are approximately 25.0 and 43.7 percentage points, respectively.
The US loses 66.9\% of trajectory diversity despite a 0.014111 decline in this share, indicating contraction across the distribution.

\begin{table*}[t]
\centering
\caption{Mean within-case change over 2026--2036 (200 trials, six bins per coordinate).
Entropy is normalized, and $\Delta N_{\mathrm{eff}}$ reports percentage change in trajectory diversity.}
\label{tab:scenario-change}
\small
\setlength{\tabcolsep}{3pt}
\begin{tabular*}{\textwidth}{@{\extracolsep{\fill}}lrrrrr@{}}
\toprule
Case & $\Delta q_{\mathrm{top}}$ & $\Delta H_{\mathrm{norm}}$ &
$\Delta\overline C$ & $\Delta W$ & $\Delta N_{\mathrm{eff}}$ (\%) \\
\midrule
China  & $+0.014307$ & $-0.174928$ & $+0.076772$ & $+0.148896$ & $-71.5$ \\
Russia       & $+0.250273$ & $-0.133861$ & $+0.129183$ & $-0.040939$ & $-61.6$ \\
Japan        & $+0.436691$ & $-0.217541$ & $+0.112458$ & $-0.073577$ & $-79.0$ \\
UK  & $+0.044018$ & $-0.103967$ & $+0.090929$ & $+0.097450$ & $-52.6$ \\
US & $-0.014111$ & $-0.152817$ & $+0.067269$ & $+0.157979$ & $-66.9$ \\
\bottomrule
\end{tabular*}
\end{table*}

Cost burdens increase in all five cases.
Welfare, interpreted only as a within-case temporal change in the implemented index, decreases in Russia and Japan but increases in China, UK, and US.
Under the welfare definition, the combined contribution of activity and support in China, the UK, and the US more than offsets the increased cost penalty.

\subsection{Uncertainty and Robustness}

For each case, we first average the five models' changes within each trial, then bootstrap the resulting 200 trial-level ensemble changes to form 95\% intervals.
All 25 change intervals exclude zero, with diversity intervals computed for absolute changes and Table~\ref{tab:scenario-change} reporting percentage changes.
The top-share interval is $[0.435086,0.438237]$ in Japan and $[-0.014849,-0.013344]$ in the US, supporting their contrasting top-share directions.
These intervals summarize variation across simulation trials.

Trajectory collapse persists with four, six, or eight bins per activity coordinate: normalized entropy and trajectory diversity decline in all 15 case--partition combinations.
Parameter robustness is assessed using 64 quasi-random draws around each fitted setting and 20 simulation trials per draw.
Every tested perturbation preserves declining diversity and entropy, rising costs, and the within-case welfare direction.
Top share increases in 99.7\%, 100\%, 100\%, 99.8\%, and 2.8\% of perturbations, respectively, in the order used in Table~\ref{tab:scenario-change}.

Increasing the number of simulated agents tenfold so that one agent represents approximately 10,000 people preserves every qualitative direction across all five cases and all three partitions.
At six bins, normalized-entropy changes range from $-0.152$ to $-0.226$, and trajectory diversity declines by 66.4\% to 80.3\%.
For Japan, 200 simulation trials for each of three ten-year windows give similar directions and magnitudes across initialization years (Table~\ref{tab:tokyo-fixed-horizon}).

\begin{table}[t]
\centering
\caption{Japan sensitivity to initialization year over ten-year windows (200 trials each).
Entropy is normalized, and $\Delta N_{\mathrm{eff}}$ is an absolute change.}
\label{tab:tokyo-fixed-horizon}
\small
\setlength{\tabcolsep}{3pt}
\begin{tabular*}{\textwidth}{@{\extracolsep{\fill}}lrrrrr@{}}
\toprule
Window & $\Delta q_{\mathrm{top}}$ & $\Delta H_{\mathrm{norm}}$ &
$\Delta\overline C$ & $\Delta W$ & $\Delta N_{\mathrm{eff}}$ \\
\midrule
2010--2020 & $+0.425141$ & $-0.212091$ & $+0.112891$ & $-0.068483$ & $-9.04297$ \\
2024--2034 & $+0.438000$ & $-0.217996$ & $+0.112435$ & $-0.072877$ & $-8.82712$ \\
2025--2035 & $+0.437407$ & $-0.217861$ & $+0.112554$ & $-0.073363$ & $-8.80245$ \\
\bottomrule
\end{tabular*}
\end{table}

\subsection{Structural Perturbations and Transfer}

Nine connectivity--visibility perturbations, evaluated with 100 simulation trials, yield similar population fits.
All beat persistence except in China, where the Hellinger range of 0.050013--0.050318 straddles its persistence distance of 0.050175.
Cross-case transfer uses the final ensembles and 100 simulation trials (Table~\ref{tab:cross-case-transfer}).
The own-case ensemble fits best in all five targets, but off-case degradation varies from 8\% to 86\%.
The smallest transfer penalty occurs in the US and the largest in the UK, showing that the sensitivity of population fit to transferred settings varies substantially across cases.
The shared model family therefore requires fitting to each urban system's population distribution and regional configuration.

\begin{table}[t]
\centering
\caption{Own-case and best transferred ensemble Hellinger distances (100 trials).
The source identifies the best off-case ensemble.}
\label{tab:cross-case-transfer}
\small
\begin{tabular*}{\textwidth}{@{\extracolsep{\fill}}lrrl@{}}
\toprule
Target & Own case & Best off case & Off-case source \\
\midrule
China  & 0.050244 & 0.055896 & UK \\
Russia       & 0.022327 & 0.028396 & Japan \\
Japan        & 0.014021 & 0.019020 & Russia \\
UK  & 0.013338 & 0.024856 & China \\
US & 0.052301 & 0.056616 & China \\
\bottomrule
\end{tabular*}
\end{table}

\section{Discussion}

The US result shows why concentration should be assessed across the full activity distribution.
Russia and Japan combine diversity loss with growth in the most common trajectory, whereas US diversity declines alongside its top share.
Rising costs accompany both improving and declining welfare across the cases, supporting the joint use of diversity, cost, and welfare in regional-development analysis.
The model provides a basis for future comparisons of public-institution relocation, regional support, and transport investment.
Such comparisons should assess activity diversity and cost burdens alongside within-case welfare.
For relocation scenarios, maintaining population in secondary regions and sustaining diverse activity orientations would constitute distinct outcomes.

Extrapolating the model fitted to the 2010--2026 population mapping to 2036 requires further validation.
The 2020--2025 comparison checks population shares within the fitting interval, without directly validating simulated activity orientations.
Observed mobility and accessibility data could provide direct checks on the simulated activity orientations.
Independent-year validation and comparisons with fitted spatial-interaction baselines remain for future work.
Differences in regional coverage limit cross-case comparisons, and bootstrap intervals cover only variation across simulation trials.
Further evaluation should address data and model uncertainty and isolate the effects of topology, connectivity, and visibility.

\section{Conclusion}

We use an agent-based model fitted to the 2010--2026 population mapping to examine trajectory diversity in 2036 scenarios across five country cases.
Diversity declines by 52.6--79.0\%, and cost burdens rise in every case.
The Russian and Japanese cases show the strongest increases in the share of the most common trajectory, while the US case loses diversity despite a slight decline in this share.
Welfare changes differ across cases, showing that reduced activity diversity can accompany either improving or declining welfare.
These findings suggest that regional development should be assessed through activity diversity, cost burdens, and welfare together.

\paragraph{Acknowledgments}{
    This research was supported by the AI Seoul Tech Research Support Program of the Seoul Future Foundation.
}

\bibliographystyle{splncs04}
\bibliography{references}

\end{document}